\documentclass [a4paper, 11pt] {article}

\usepackage{pos}

 \usepackage{epsfig}
 \newcommand{\rmF}  {\sp{19}  {\rm F}}
 \newcommand{\rmXe} {\sp{129} {\rm Xe}}
 \def \PeriodCa     { 19.49 --  79.49}
 \def \PeriodCb     {110.74 -- 170.74}
 \def \PeriodCc     {201.99 -- 261.99}
 \def \PeriodCd     {293.24 -- 353.24}
 \def \PlotNumberCa {04949}
 \def \PlotNumberCb {14074}
 \def \PlotNumberCc {23199}
 \def \PlotNumberCd {32324}
\newcommand{\OnlinePlotfveffGmchi} [1] { 
\href{http://www.tir.tw/phys/hep/dm/amidas-2d/amidas-2d.php%
      ?amidas_2D_function=fv_eff%    #1%
      &mode_fv_eff=N_v%              #2=#3%
      &frame=G%                      #4%
      &mode_animation=mchi%          #5%
      &target=\Target%               #6%
      &period=periodA%               #6%
      &event_No=500}            
     {#1}
}
\newcommand{\OnlinePlotfveffangGmchi} [2] { 
\href{http://www.tir.tw/phys/hep/dm/amidas-2d/amidas-2d.php%
      ?amidas_2D_function=fv_eff%    #1%
      &mode_fv_eff=#1_ang%           #2=#3%
      &frame=G%                      #4%
      &mode_animation=mchi%          #5%
      &target=\Target%               #6%
      &period=periodA%               #6%
      &event_No=500}            
     {#2}
}
\newcommand{\OnlinePlotfveffGAnnual} [1] { 
\href{http://www.tir.tw/phys/hep/dm/amidas-2d/amidas-2d.php%
      ?amidas_2D_function=fv_eff%          #1%
      &mode_fv_eff=N_v%                    #2=#3%
      &frame=G%                            #4%
      &mode_animation=annual%              #5%
      &target=\Target%                     #6%
      &mchi=200%                    #1%    #7%
      &period=periodC%                     #6%
      &event_No=500}                  
     {#1}
}
\newcommand{\OnlinePlotfveffangGAnnual} [2] { 
\href{http://www.tir.tw/phys/hep/dm/amidas-2d/amidas-2d.php%
      ?amidas_2D_function=fv_eff%         #1%
      &mode_fv_eff=#1_ang%                #2=#3%
      &frame=G%                           #4%
      &mode_animation=annual%             #5%
      &target=\Target%                    #6%
      &mchi=200%                    #1%   #7%
      &period=periodC%                    #6%
      &event_No=500}                  
     {#2}
}
\newcommand{\InsertFigurefveffGAnnual} [1] {
\begin{figure} [t!]
\begin{center}
   \begin{minipage} {\Subplotwidth}
    {\begin{center}
     {\footnotesize \PeriodCa\ day}
     \end{center}}
   \end{minipage}
 \hspace{0.2 cm}
   \begin{minipage} {\Subplotwidth}
    {\begin{center}
     {\footnotesize \PeriodCb\ day}
     \end{center}}
   \end{minipage}
 \hspace{0.2 cm}
   \begin{minipage} {\Subplotwidth}
    {\begin{center}
     {\footnotesize \PeriodCc\ day}
     \end{center}}
   \end{minipage}
 \hspace{0.2 cm}
   \begin{minipage} {\Subplotwidth}
    {\begin{center}
     {\footnotesize \PeriodCd\ day}
     \end{center}}
   \end{minipage}
 \\
 \OnlinePlotfveffGAnnual
  {\begin{minipage} {\Subplotwidth}
    {\begin{center}
      \includegraphics [width = \Subplotwidth]
                       {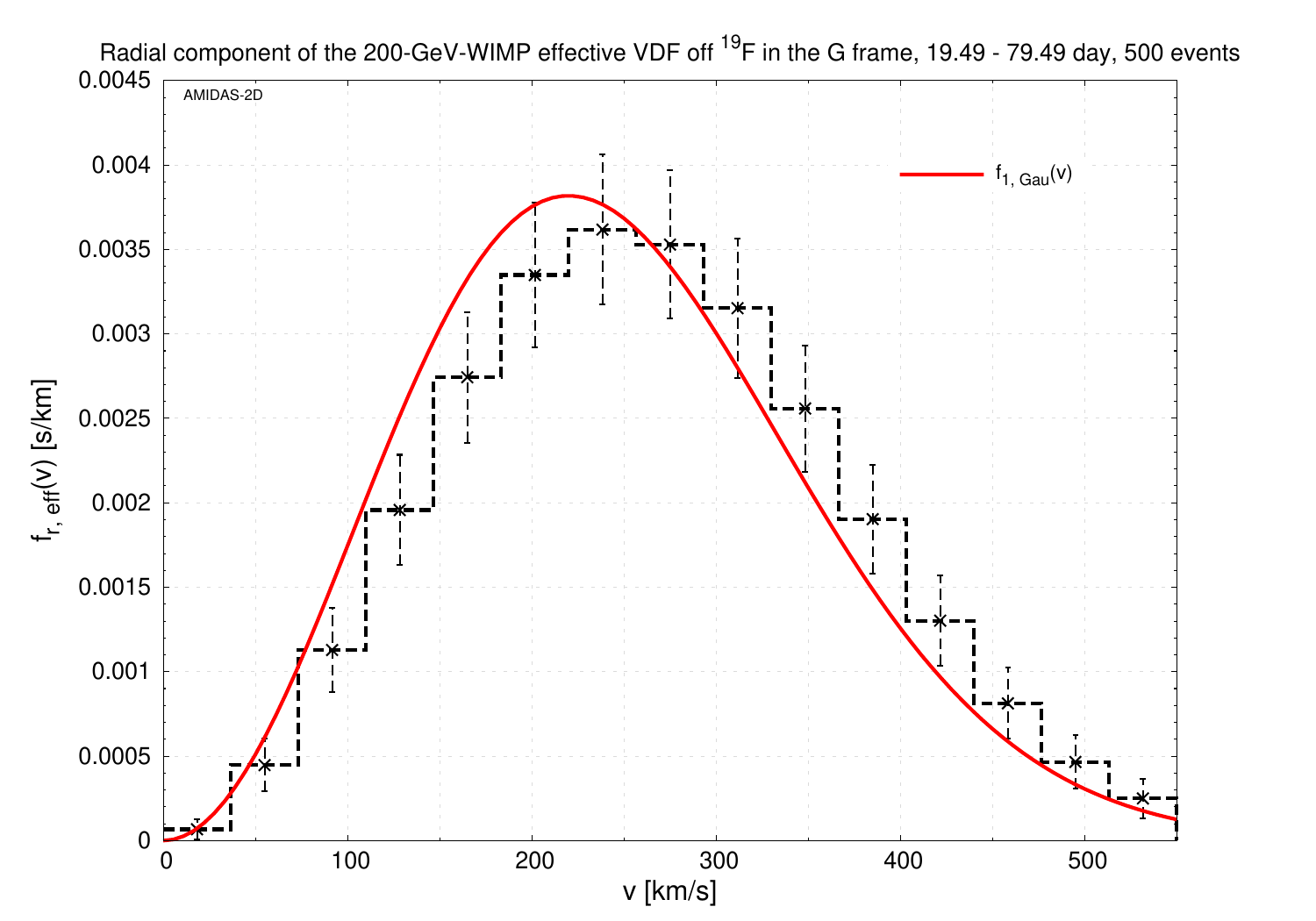}
     \end{center}}
   \end{minipage}}
 \hspace{0.1 cm}
 \OnlinePlotfveffGAnnual
  {\begin{minipage} {\Subplotwidth}
    {\begin{center}
      \includegraphics [width = \Subplotwidth]
                       {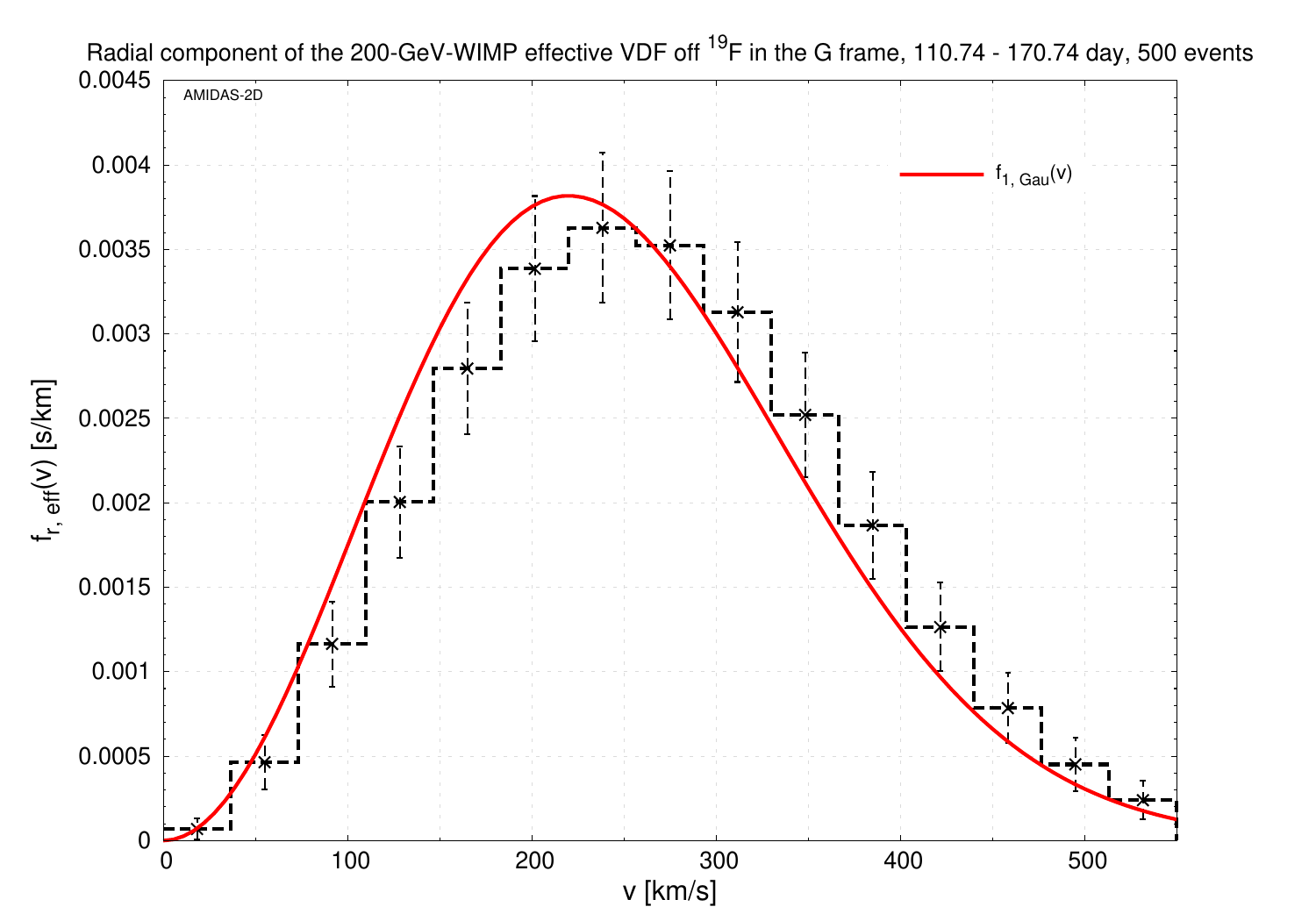}
     \end{center}}
   \end{minipage}}
 \hspace{0.1 cm}
 \OnlinePlotfveffGAnnual
  {\begin{minipage} {\Subplotwidth}
    {\begin{center}
      \includegraphics [width = \Subplotwidth]
                       {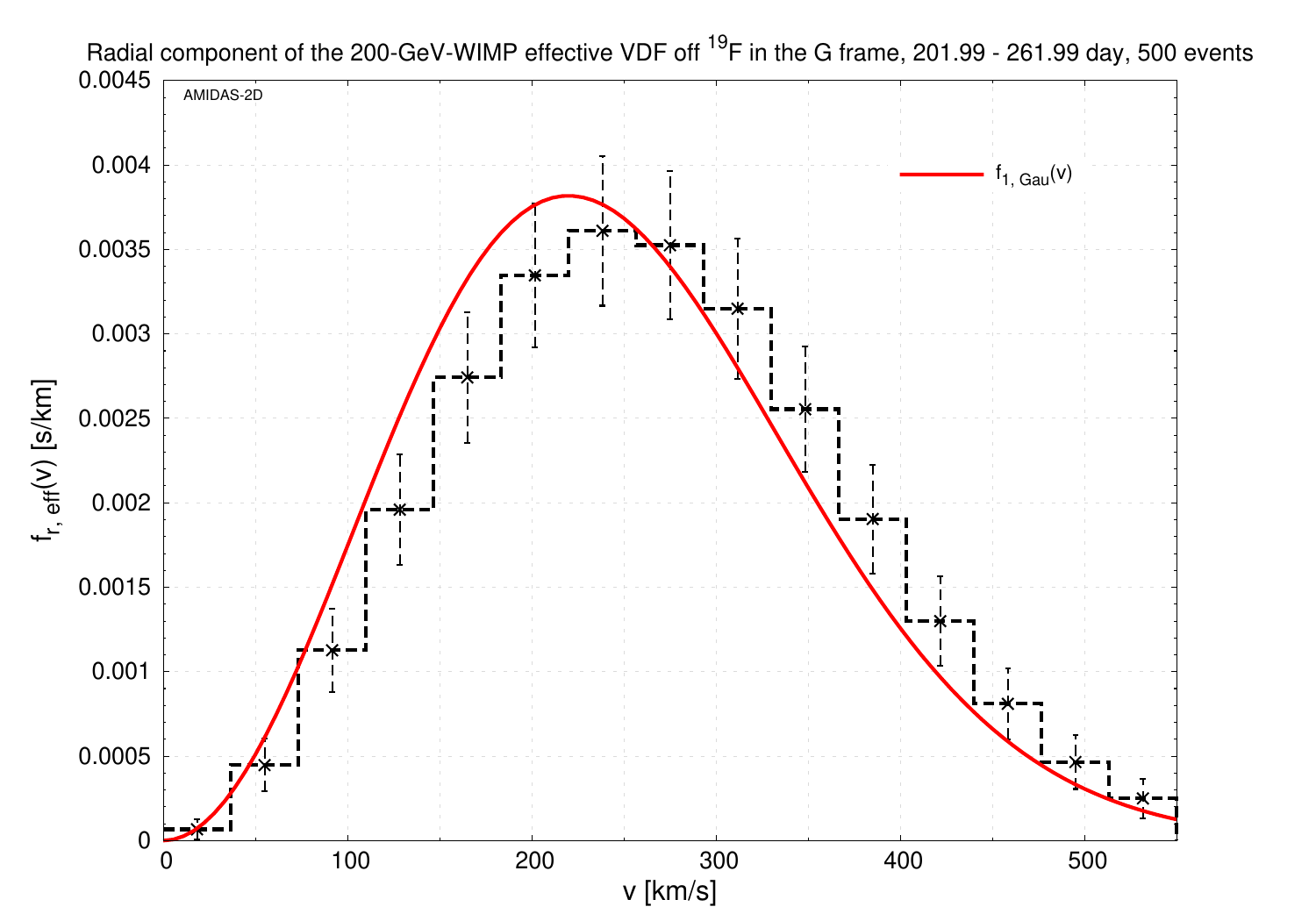}
     \end{center}}
   \end{minipage}}
 \hspace{0.1 cm}
 \OnlinePlotfveffGAnnual
  {\begin{minipage} {\Subplotwidth}
    {\begin{center}
      \includegraphics [width = \Subplotwidth]
                       {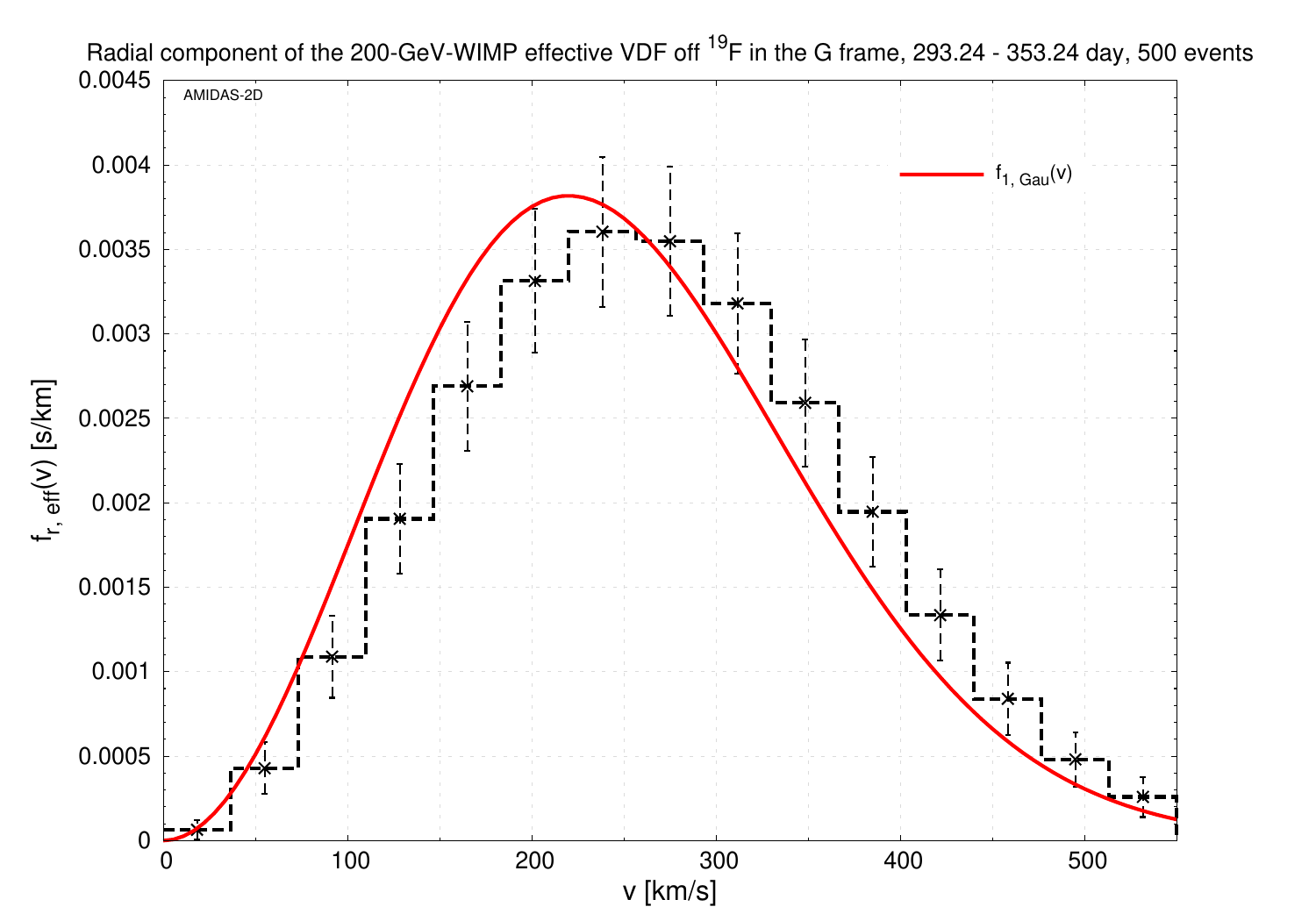}
     \end{center}}
   \end{minipage}}
 \\ \vspace{ 0.1 cm}
 (a) The radial component
 \\ \vspace{ 0.2 cm}
 \OnlinePlotfveffangGAnnual
  {N}
  {\begin{minipage} {\Subplotwidth}
    {\begin{center}
      \includegraphics [width = \Subplotwidth]
                       {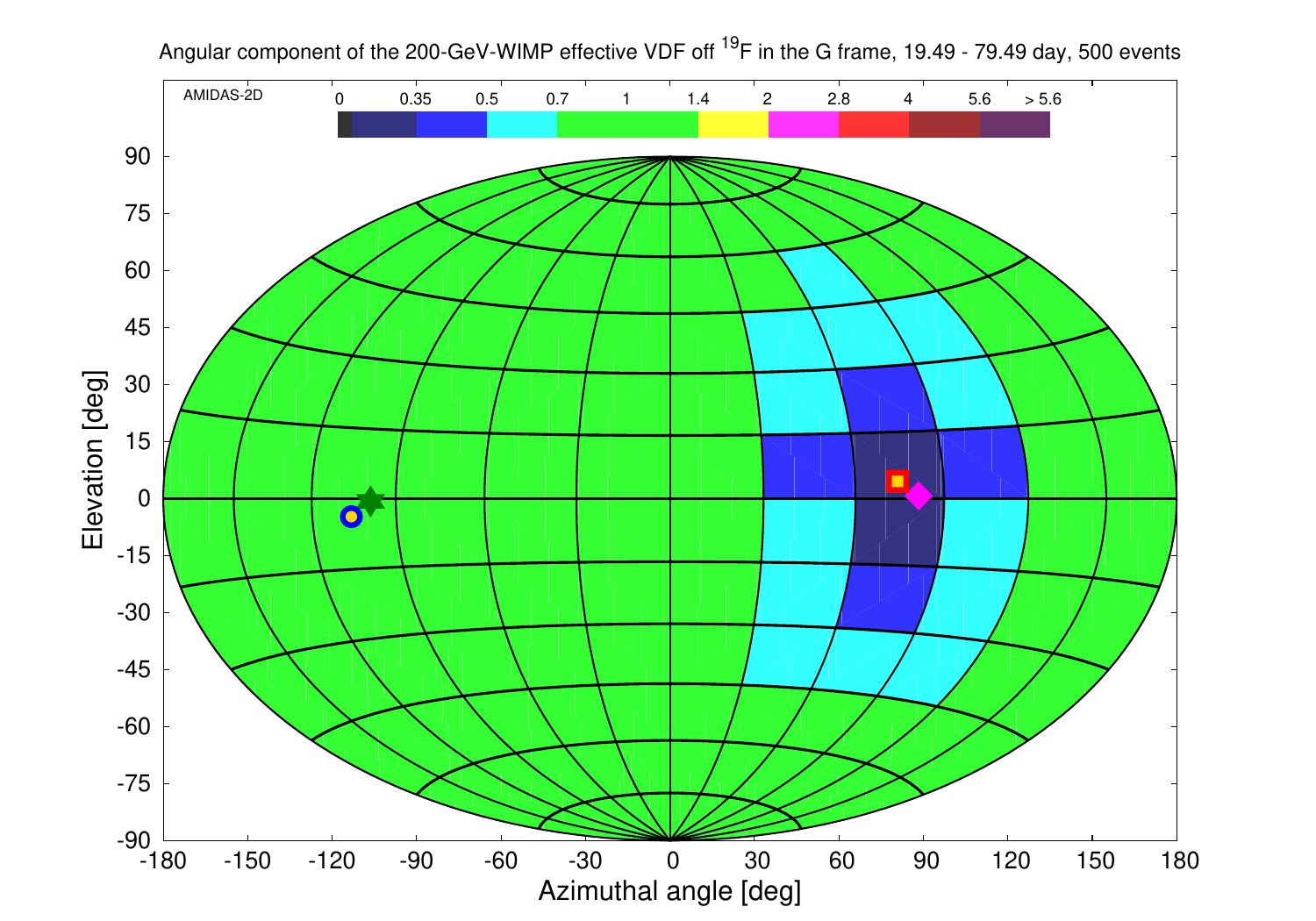}
     \end{center}}
   \end{minipage}}
 \hspace{0.1 cm}
 \OnlinePlotfveffangGAnnual
  {N}
  {\begin{minipage} {\Subplotwidth}
    {\begin{center}
      \includegraphics [width = \Subplotwidth]
                       {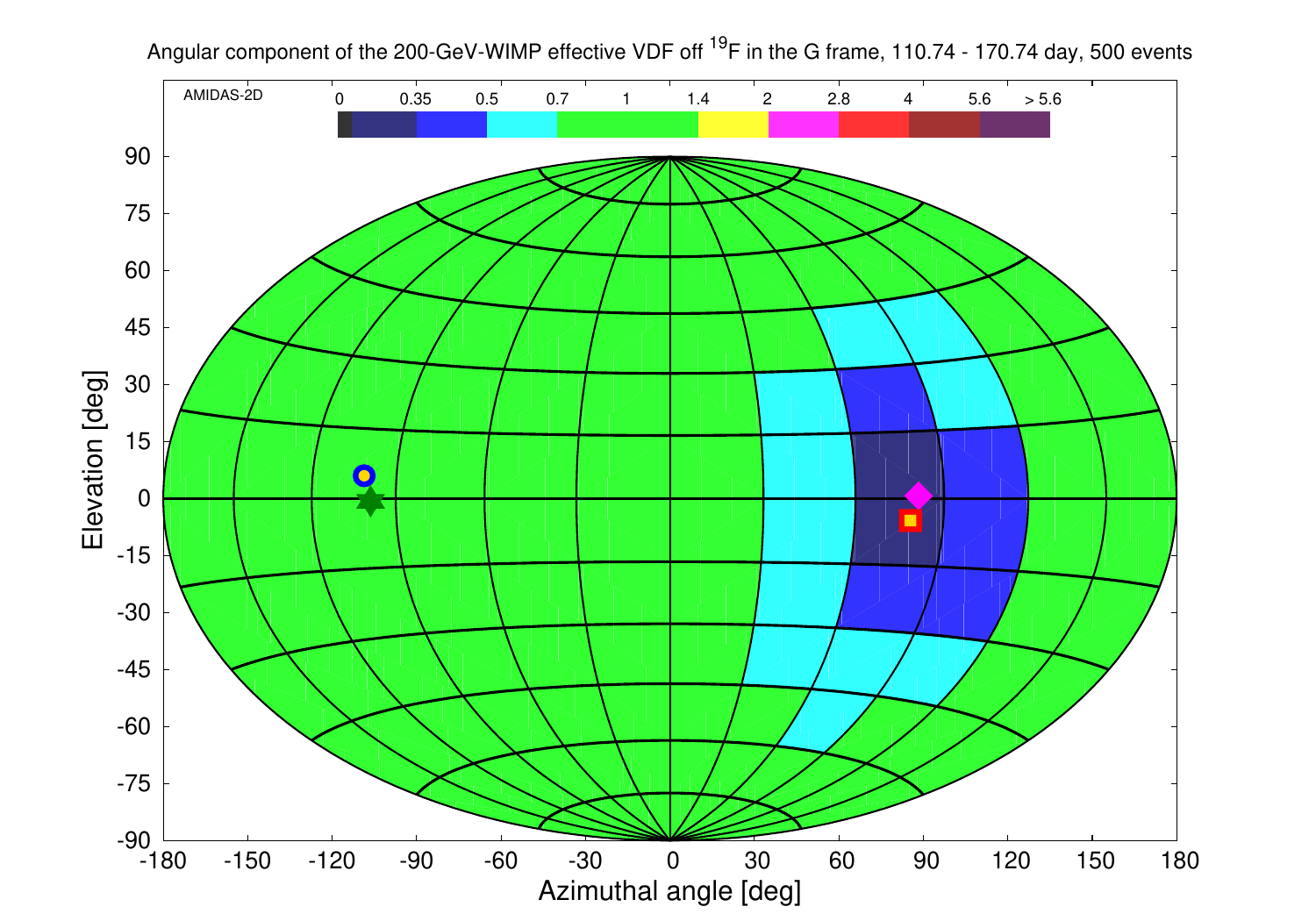}
     \end{center}}
   \end{minipage}}
 \hspace{0.1 cm}
 \OnlinePlotfveffangGAnnual
  {N}
  {\begin{minipage} {\Subplotwidth}
    {\begin{center}
      \includegraphics [width = \Subplotwidth]
                       {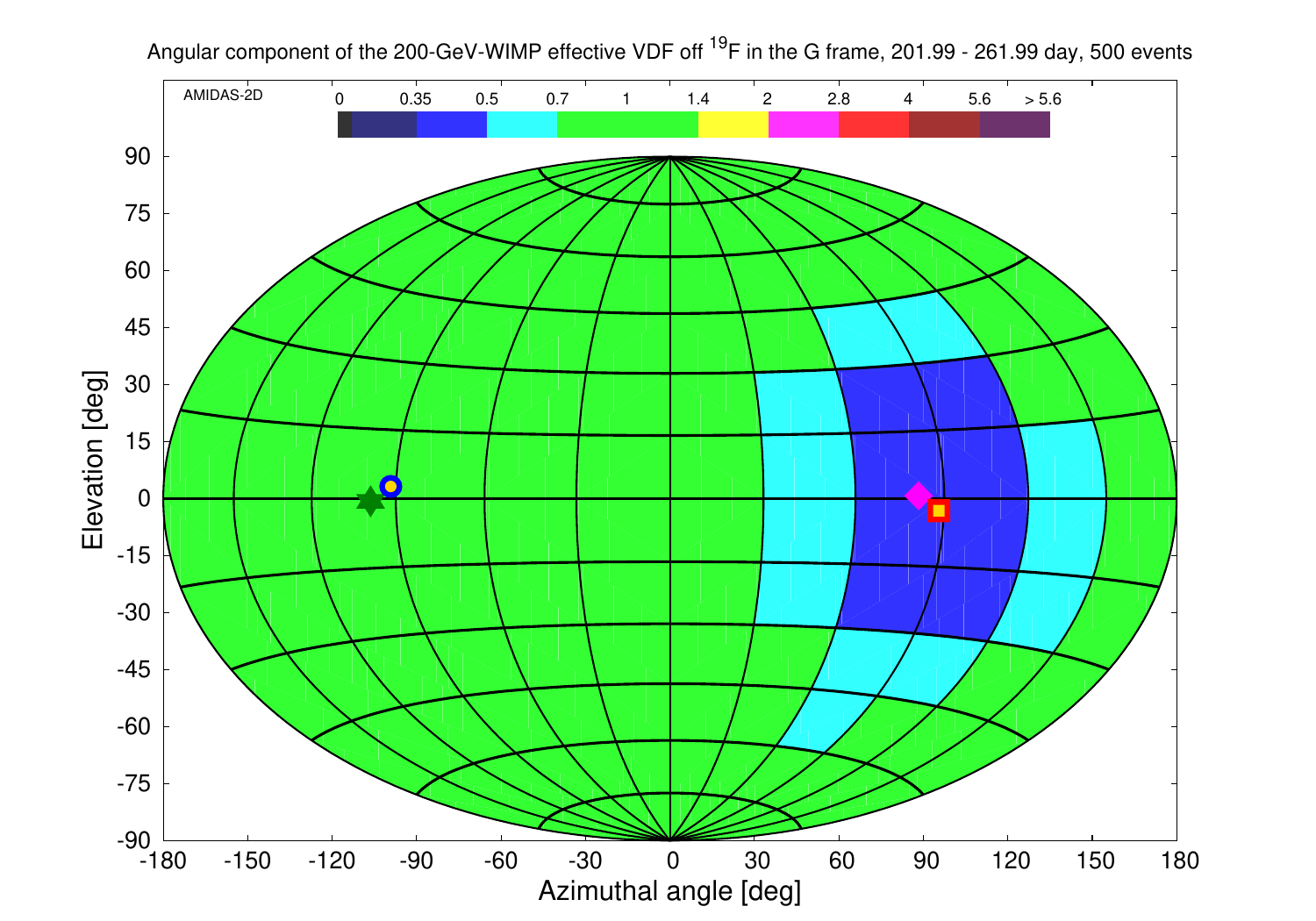}
     \end{center}}
   \end{minipage}}
 \hspace{0.1 cm}
 \OnlinePlotfveffangGAnnual
  {N}
  {\begin{minipage} {\Subplotwidth}
    {\begin{center}
      \includegraphics [width = \Subplotwidth]
                       {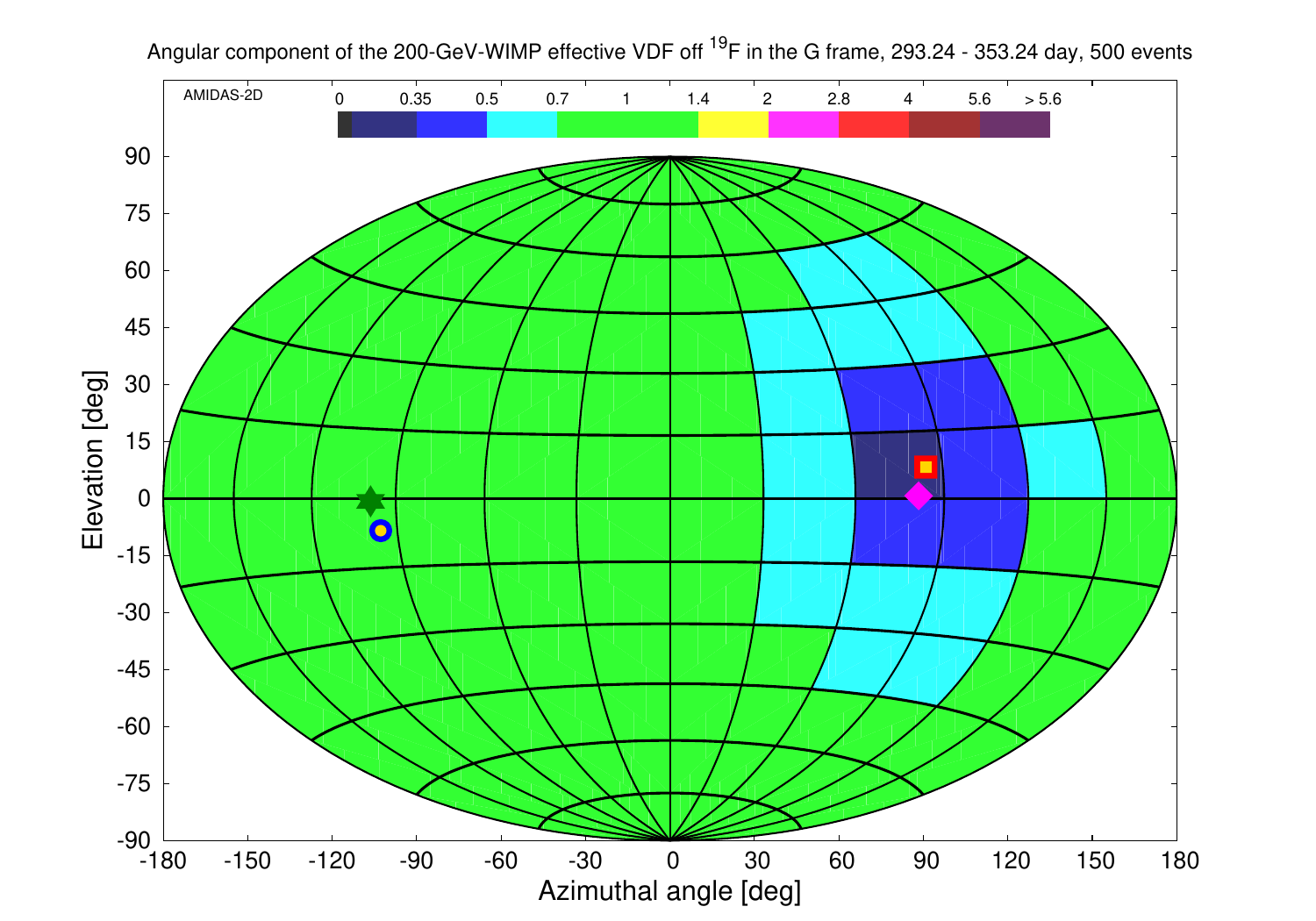}
     \end{center}}
   \end{minipage}}
 \\ \vspace{ 0.1 cm}
 (b) The angular component
 \\ \vspace{-0.4 cm}
\end{center}
\caption{
 #1
}
\label{fig:N-\Target-\WIMPmass-G-0500-04949}
\end{figure}
}

\title{3-Dimensional WIMP Effective Velocity Distribution}

\author*{Chung-Lin Shan}

\affiliation{%
 Preparatory Office of
 the Supporting Center for
 Taiwan Independent Researchers                       \\
 P.O.BOX 21 National Yang Ming Chiao Tung University,
 Hsinchu City 300093, Taiwan, R.O.C.}

\emailAdd{clshan@tir.tw}

\abstract{%
 In this talk,
 I discussed
 a 3-dimensional
 ``effective'' velocity distribution of
 Weakly Interacting Massive Particles (WIMPs),
 which,
 instead of the theoretically predicted velocity distribution of
 ``entire'' Galactic Dark Matter particles,
 describes the actual
 velocity distribution of WIMPs
 ``scattering off'' (specified) target nuclei
 in an underground detector.
 Based on
 numerical results carried out by
 our double Monte Carlo scattering--by--scattering simulation of
 3-dimensional elastic WIMP--nucleus scattering,
 an (asymmetric) ``forward--backward asymmetry''
 was also demonstrated.
}

\FullConference{%
 19th International Conference on Topics in Astroparticle and Underground Physics (TAUP2025) \\
 24–30 Aug 2025 \\
 Xichang, China \\}

\begin{document}
\maketitle

\section{Introduction}

 For most people
 working on direct Dark Matter (DM) detection physics,
 it might be assumed automatically that
 the subgroup of Weakly Interacting Massive Particles (WIMPs)
 scattering off target nuclei in an underground detector
 would have the same astronomical and kinematic properties:
 3-dimensional velocity distribution,
 average and root--mean--square velocities,
 average kinetic energy
 and so on,
 as the entire Galactic halo WIMPs
 impinging on a direct DM detector,
 but not necessarily scattering off target nuclei.

 In Ref.~\cite{DMDDD-3D-WIMP-N},
 we accomplished
 our double Monte Carlo
 scattering--by--scattering simulation package
 for the 3-D elastic WIMP--nucleus scattering process.
 For {\em each} WIMP scattering event,
 our package generates
 a 3-D velocity of the incident WIMP
 in the Galactic coordinate system
 according to a chosen generating velocity distribution
 and a scattering time
 in the observation window,
 transfers them into
 the laboratory coordinate system,
 carries out the scattering off a target nucleus
 with the recoil direction and the recoil energy
 satisfying our probability criterion
 (or abandons this case
  when
  the recoil direction doesn't pass the criterion
  and/or the recoil energy is out of the experimental measuring window).
 This workflow allows us to investigate
 properties of the WIMPs
 {\em not only} impinging on a direct DM detector,
 {\em but also} scattering off target nuclei.
 Then,
 instead of the
 theoretically predicted velocity distribution of
 the entire Galactic halo WIMPs,
 we introduced in Ref.~\cite{DMDDD-fv_eff}
 the (WIMP--mass-- and target--dependent)
 3-dimensional WIMP {\em effective} velocity distribution
 to
 describe the actual
 velocity distribution of the scattering WIMPs.

\section{3-D WIMP Galactic effective velocity distribution}
\label{sec:fv_eff-G}

 Given
 the mass of Galactic halo WIMPs,
 two factors can affect the scattering probability of these WIMPs
 moving with different incident velocities.
 Firstly,
 it is easy to understand that
 the higher the incident velocity,
 the larger the WIMP flux
 and thus the larger the scattering probability
 (in a unit time).
 Secondly,
 the higher the incident velocity
 and thus the larger the kinetic energy,
 the larger the transferable recoil energy to the scattered target nucleus
 and the stronger the nuclear form factor suppression,
 thus the smaller the scattering probability.

 \def \Subplotwidth {4 cm}
\begin{figure} [t!]
\begin{center}
 \def \Target   {F19}
 \def \WIMPmass {0200}
 \OnlinePlotfveffGmchi
  {\begin{minipage} {\Subplotwidth}
    {\begin{center}
      \includegraphics [width = \Subplotwidth]
                       {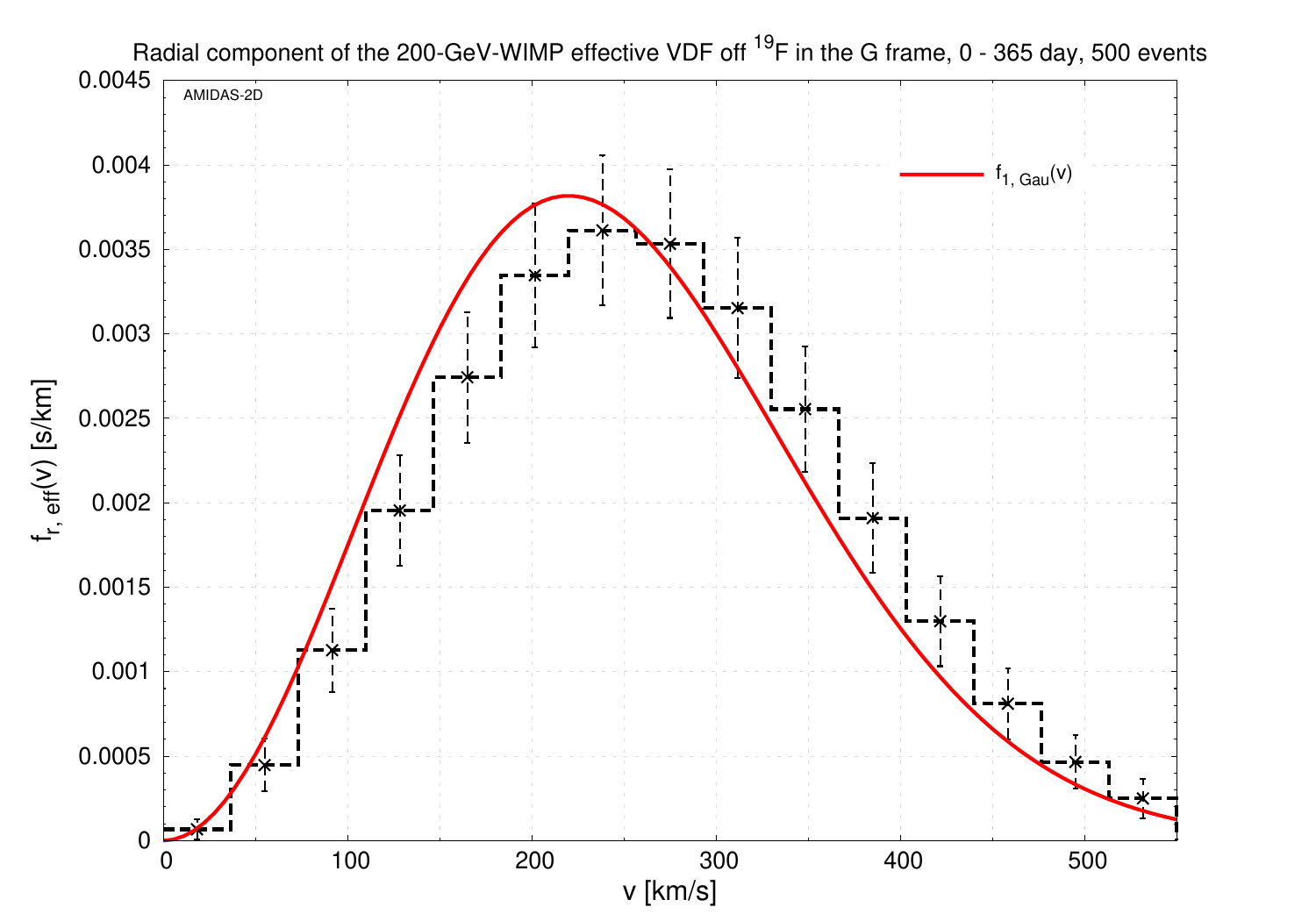}
     \end{center}}
   \end{minipage}}
 \hspace{0.2 cm}
 \OnlinePlotfveffangGmchi
  {N}
  {\begin{minipage} {\Subplotwidth}
    {\begin{center}
      \includegraphics [width = \Subplotwidth]
                       {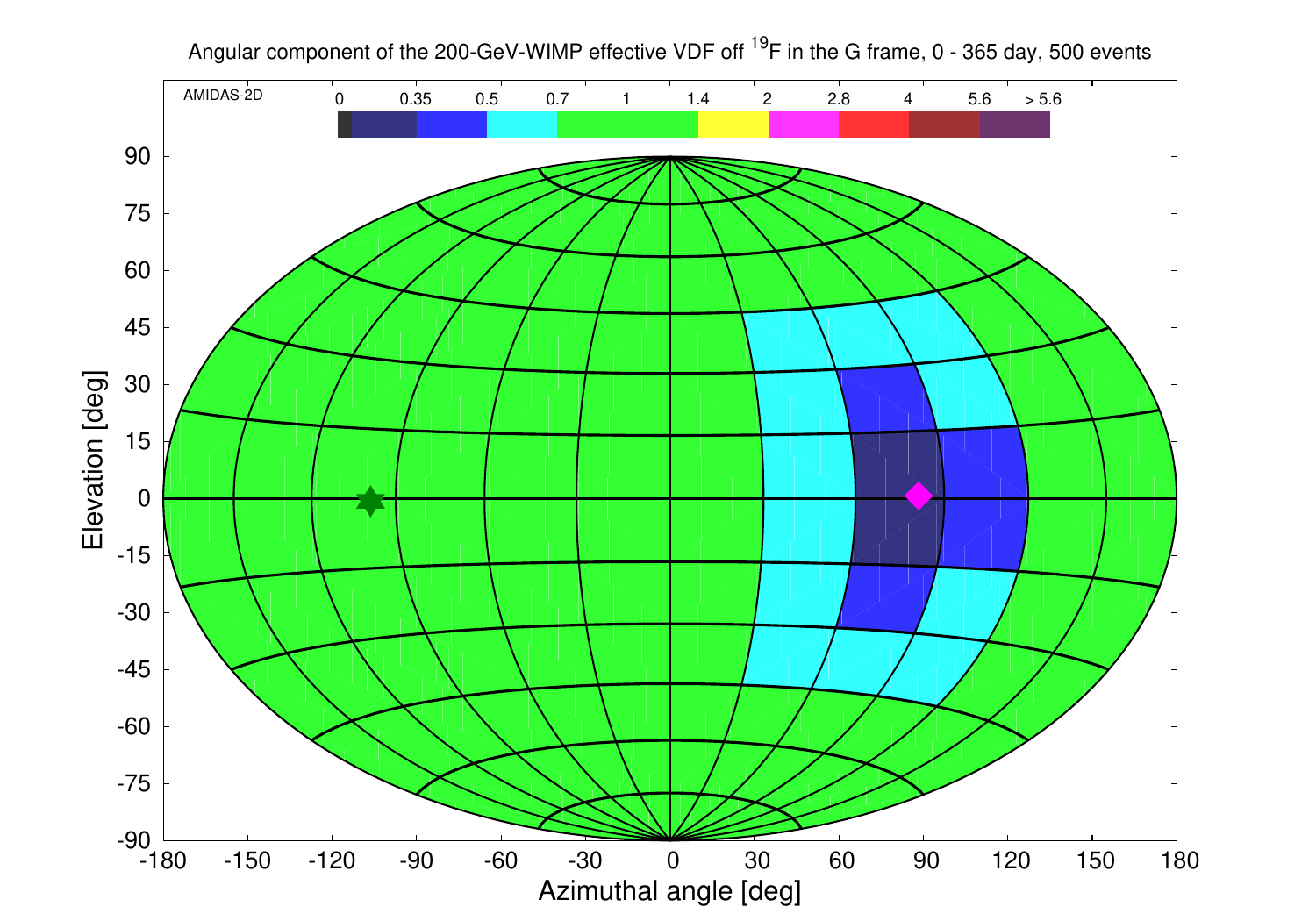}
     \end{center}}
   \end{minipage}}
 \hspace{0.2 cm}
 \OnlinePlotfveffangGmchi
  {PoN}
  {\begin{minipage} {\Subplotwidth}
    {\begin{center}
      \includegraphics [width = \Subplotwidth]
                       {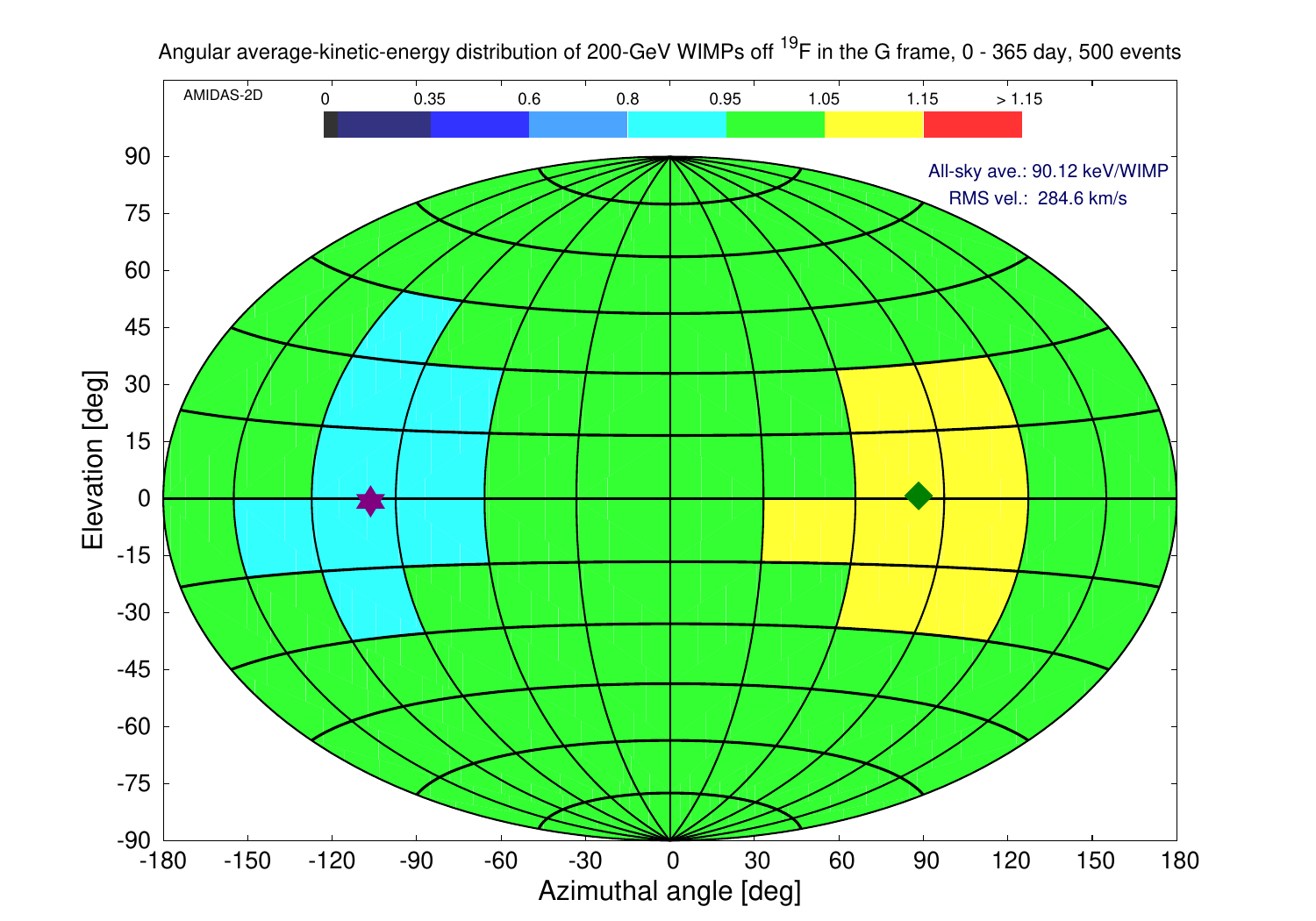}
     \end{center}}
   \end{minipage}}
 \\ \vspace{ 0.1 cm}
 (a) 200-GeV WIMPs scatter off $\rmF$
 \\ \vspace{ 0.2 cm}
 \def \Target   {Xe129}
 \OnlinePlotfveffGmchi
  {\begin{minipage} {\Subplotwidth}
    {\begin{center}
      \includegraphics [width = \Subplotwidth]
                       {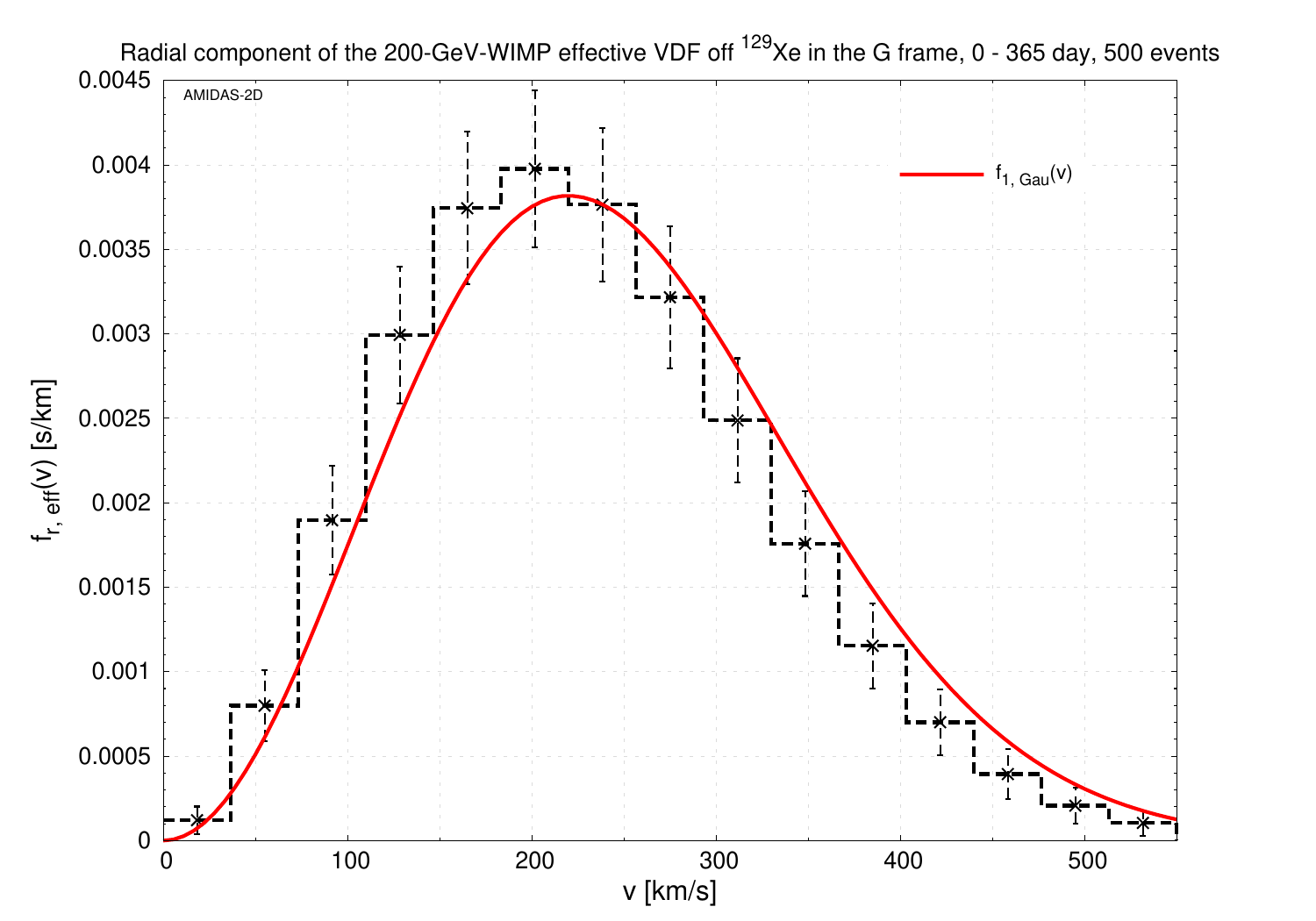}
     \end{center}}
   \end{minipage}}
 \hspace{0.2 cm}
 \OnlinePlotfveffangGmchi
  {N}
  {\begin{minipage} {\Subplotwidth}
    {\begin{center}
      \includegraphics [width = \Subplotwidth]
                       {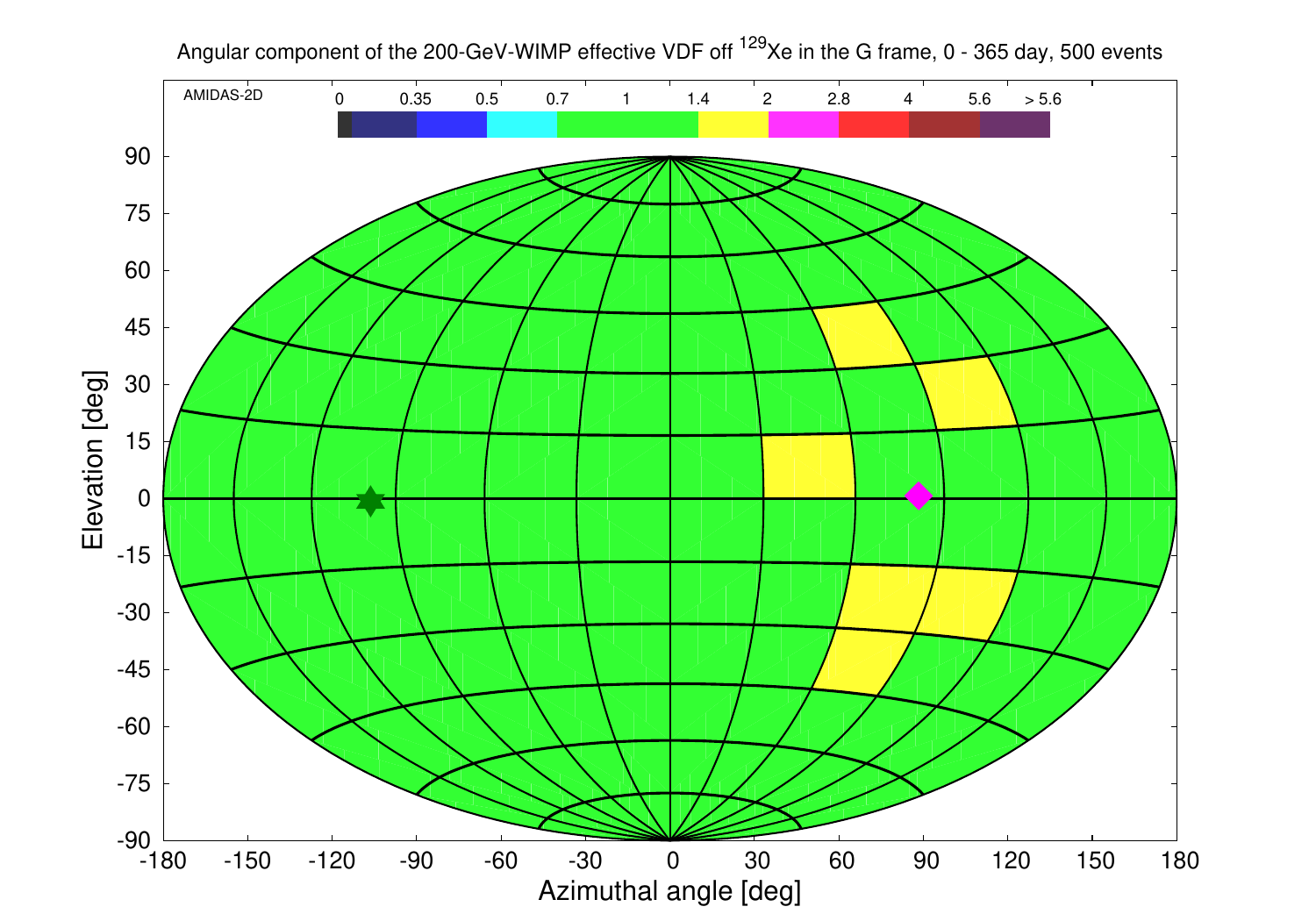}
     \end{center}}
   \end{minipage}}
 \hspace{0.2 cm}
 \OnlinePlotfveffangGmchi
  {PoN}
  {\begin{minipage} {\Subplotwidth}
    {\begin{center}
      \includegraphics [width = \Subplotwidth]
                       {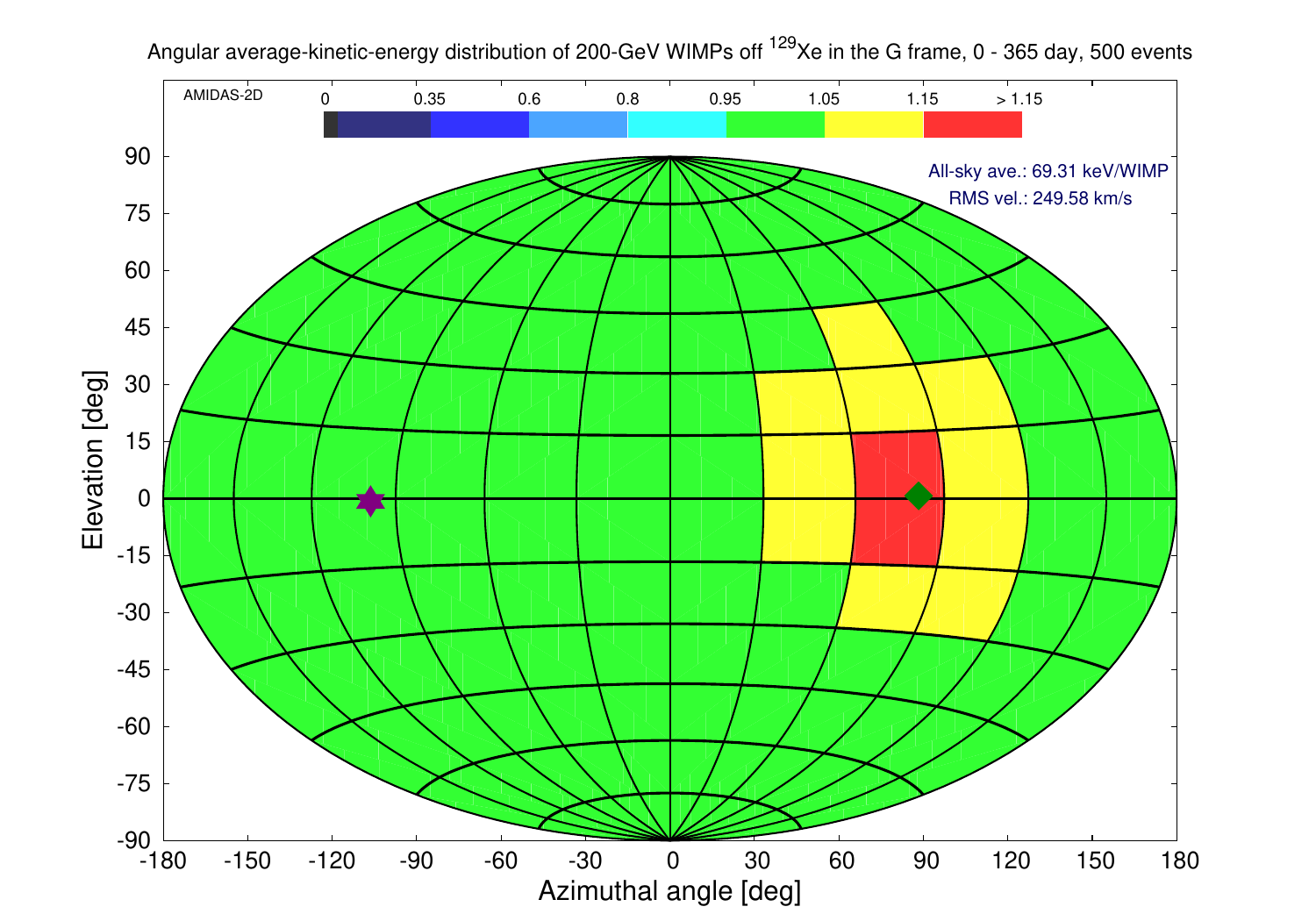}
     \end{center}}
   \end{minipage}}
 \\ \vspace{ 0.1 cm}
 (b) 200-GeV WIMPs scatter off $\rmXe$
 \\ \vspace{-0.4 cm}
\end{center}
\caption{
 The radial (left)
 and the angular (middle) components
 as well as
 the average kinetic energy (right) of
 the 3-D WIMP Galactic effective velocity distribution.
 See the text for further details.
}
\label{fig:N-G-0500-00000}
\end{figure}

 In Figs.~\ref{fig:N-G-0500-00000},
 we show
 the radial (left)
 and the angular (middle) components
 as well as
 the average kinetic energy (right) of
 the 3-D WIMP effective velocity distribution
 in the Galactic coordinate system.
 $\rmF$ and $\rmXe$
 have been considered as
 a light and a heavy target nucleus
 scattered by 200-GeV WIMPs.
 5,000 experiments
 with
 500 total events on average
 (Poisson--distributed)
 in one experiment
 in one entire year
 have been simulated
 and binned into
 15 bins for the radial component
 and 12 $\times$ 12 bins
 for the angular (longitude and latitude) component,
 respectively.
 The solid red curves
 in the left plots
 are the generating simple Maxwellian velocity distribution
 \cite{DMDDD-3D-WIMP-N}.
 In the middle and right plots,
 while
 the magenta/dark--green diamonds
 on the right--hand (eastern) sky indicate
 the moving direction of the Solar system
 in the Galactic coordinate system
 \cite{DMDDD-N, DMDDD-3D-WIMP-N}:
 0.60$^{\circ}$N, 81.22$^{\circ}$E,
 the dark--green/purple stars
 on the left--hand (western) sky indicate
 the theoretical main direction of incident WIMPs
 \cite{DMDDD-N, DMDDD-3D-WIMP-N}:
 0.60$^{\circ}$S, 98.78$^{\circ}$W.

 One can easily find that
\begin{enumerate}
\setlength{\itemsep}{-0.1 cm}
\item
 The average and the root--mean--square velocities
 (in turn the average kinetic energy) of
 the incident halo WIMPs
 scattering off
 light (heavy) target nuclei
 like $\rmF$ ($\rmXe$)
 are higher (lower) than
 those of
 the entire Galactic (incident) halo WIMPs.
\item
 WIMPs moving (approximately)
 in the {\em same} direction as
 the Galactic movement of our Solar system
 (the laboratory/detector)
 would have (much) lower probabilities
 to scatter off light ($\rmF$) target nuclei
 than WIMPs moving (approximately)
 in the {\em opposite} direction.
\item
 But,
 once
 WIMPs are heavy (enough),
 the forward--moving and scattering WIMPs
 could have higher probabilities
 to scatter off heavy ($\rmXe$) target nuclei.
\item
 Nevertheless,
 for both of light ($\rmF$) and heavy ($\rmXe$) target nuclei,
 the forward--moving and scattering WIMPs
 would {\em always} have {\em larger}
 average and root--mean--square velocities
 (average kinetic energy)
 than the backward--moving WIMPs.
\end{enumerate}
\subsubsection*{Forward--backward asymmetry}
\label{sec:fv_eff-G-FBA}

 According to
 the simulation results shown
 in Figs.~\ref{fig:N-G-0500-00000}
 (and more)
 as well as
 the observations,
 we introduced in Ref.~\cite{DMDDD-fv_eff}
 a ``forward--backward asymmetry'' of
 the 3-D WIMP Galactic effective velocity distribution,
 which describes
 the {\em anisotropy} of
 the angular component of
 the 3-D effective velocity distribution.
 As discussed above,
 this is
 the superposition of
 the flux proportionality
 to the WIMP incident velocity
 and
 the scattering cross section (nuclear form factor) suppression:
\vspace{-0.2 cm}
\begin{enumerate}
\setlength{\itemsep}{-0.1 cm}
\item
 Generally speaking,
 the forward--moving WIMPs
 have relatively lower relative velocities
 and in turn smaller kinetic
 and transferable recoil energies.
 Thus,
 while
 the WIMP fluxes are low,
 the cross section (nuclear form factor) suppression
 on the scattering probability of these WIMPs
 could be weak or even negligible.
\item
 In contrast,
 the backward--moving WIMPs
 would hit our detector from the front
 with (much) higher incident velocities
 and in turn
  (much) larger kinetic and transferable recoil energies.
 So
 the WIMP fluxes are (much) higher,
 but
 the cross section (nuclear form factor) suppression
 could be (much) stronger.
\item
 Since
 the nuclear form factor suppression of
 light ($\rmF$) target nuclei
 is pretty week
 at even large energy,
 the flux proportionality
 to the WIMP incident velocity
 dominates
 and the forward--moving WIMPs
 would thus have smaller scattering probabilities.
\item
 Due to
 the strong nuclear form factor suppression of
 heavy ($\rmXe$) target nuclei,
 the forward--moving WIMPs
 would then have larger scattering probabilities.
\end{enumerate}

 It would also be interesting to mention that
 the forward--backward asymmetry of
 the WIMP effective velocity distribution
 would itself be ``asymmetric'':
 the increases or decreases of
 the incident flux/(average) kinetic energy of
 the forward--moving WIMPs
 are (much) stronger.
 This would be due to
 the Galactic movement of our Solar system
 (the laboratory/detector)%
\footnote{
 Scenarios with different Solar Galactic orbital speed
 and/or a bulk rotation of the Dark Matter halo
 is currently under investigation.
}.
\subsubsection*{Annual modulation}
\label{sec:annual}
 \def \Subplotwidth {3.5 cm}
 \def \Target   {F19}
 \def \WIMPmass {0200}
 \InsertFigurefveffGAnnual
  {The radial (a)
   and the angular (b) components of
   the 3-D WIMP Galactic effective velocity distribution
   simulated with 500 total events on average
   in each 60-day observation period of four {\em advanced} seasons
   \cite{DMDDD-N, DMDDD-3D-WIMP-N}.
   200-GeV WIMPs scatter off $\rmF$.
   See the text for further details.
   }
 \def \Target   {Xe129}
 \InsertFigurefveffGAnnual
  {The same as Figs.~\ref{fig:N-F19-0200-G-0500-04949},
   except that
   $\rmXe$ is considered as our target.
   }

 In Figs.~\ref{fig:N-F19-0200-G-0500-04949}
 and \ref{fig:N-Xe129-0200-G-0500-04949},
 we show
 the radial (upper)
 and the angular (lower) components of
 the 3-D WIMP Galactic effective velocity distribution
 simulated with 500 total events on average
 in each 60-day observation period of four {\em advanced} seasons
 \cite{DMDDD-N, DMDDD-3D-WIMP-N}.
 200-GeV WIMPs scatter off the $\rmF$ and $\rmXe$ target nuclei,
 respectively.
 In each plot
 in the lower frames,
 the red/blue--yellow square
 and the blue/red--yellow circle
 indicate
 the (opposite) direction of
 the Earth's movement in the Dark Matter halo
 on the central date of the observation period
 \cite{DMDDD-N},
 respectively.

 Not surprisingly,
 we can observe
 a WIMP--mass-- and target--dependent annual modulation of
 the 3-D effective velocity distribution.
 For a WIMP mass of ${\cal O}(200)$ GeV,
\vspace{-0.2 cm}
\begin{enumerate}
\setlength{\itemsep}{-0.1 cm}
\item
 the peak of
 the radial component (magnitude) of
 the 3-D effective velocity distribution of
 the scattering WIMPs
 off light (heavy) target nuclei like $\rmF$ ($\rmXe$)
 moves towards to a lower (higher) velocity in summer
 and a higher (lower) velocity in winter
 \cite{DMDDD-fv_eff};
\item
 more clearly,
 the angular distribution pattern
 (the forward--backward asymmetry)
 on the eastern (western) sky rotates
 {\em counterclockwise} ({\em clockwise})
 following
 the instantaneous Earth's moving direction
 (theoretical main direction of incident WIMPs)
 in the Dark Matter halo
 \cite{DMDDD-fv_eff}.
\end{enumerate}
\section{Conclusions}

 In this article,
 we have demonstrated and discussed
 (the asymmetry of)
 the 3-D WIMP Galactic effective velocity distribution.
 Hopefully,
 this interesting observation
 could inspire our colleagues
 working on
 (directional) direct Dark Matter detection physics
 some novel ideas.

\subsubsection*{Acknowledgments}

 The author would like to thank
 the pleasant atmosphere of
 the W101 Ward and the Cancer Center of
 the Kaohsiung Veterans General Hospital,
 where part of this work was completed.
 This work
 was strongly encouraged by
 the ``{\it Researchers working on
 e.g.~exploring the Universe or landing on the Moon
 should not stay here but go abroad.}'' speech.

\end{document}